\documentclass[fleqn,10pt]{wlscirep}
\usepackage[utf8]{inputenc}
\usepackage[T1]{fontenc}
\usepackage{amssymb}
\usepackage{amsmath}
\usepackage{txfonts}
\usepackage{mathdots}

\title{Topological synchronization of chaotic systems}

\author[1,*]{Nir Lahav}
\author[2,3]{Irene Sendi\~na-Nadal}
\author[4]{Chittaranjan Hens}
\author[5]{Baruch Ksherim}
\author[6,7]{Baruch Barzel}
\author[8]{Reuven Cohen}
\author[9,10,11]{Stefano Boccaletti}
\affil[1]{Department of Physics, Bar-Ilan University, 52900 Ramat Gan, Israel}
\affil[2]{Complex Systems Group {\& GISC}, Universidad  Rey Juan Carlos, 28933 M\'ostoles, Madrid, Spain}
\affil[3]{Center for Biomedical Technology, Universidad Polit\'ecnica de Madrid, 28223 Pozuelo de Alarc\'on, Madrid, Spain}
\affil[4]{Physics and Applied Mathematics Unit, Indian Statistical Institute, Kolkata 700108, India}
\affil[5]{Department of Mathematics, Bar-Ilan University, 52900 Ramat Gan, Israel}
\affil[6]{Department of Mathematics, Bar-Ilan University, 52900 Ramat Gan, Israel}
\affil[7]{The Gonda Multidisciplinary Brain Research Center, Bar-Ilan University, Ramat-Gan, Israel}
\affil[8]{Department of Mathematics, Bar-Ilan University, 52900 Ramat Gan, Israel}
\affil[9]{CNR-Institute of complex systems, Via Madonna del Piano 10, 50019 Sesto Fiorentino, Italy}
\affil[10]{Moscow Institute of Physics and Technology, 9 Institutskiy per., Moscow Region 141701, Russia}
\affil[1]{Universidad Rey Juan Carlos, 28933 M\'ostoles, Madrid, Spain}
\affil[*]{\textbf{freenl@gmail.com}}

\begin{abstract}

A chaotic dynamics is typically characterized by the emergence of strange attractors with their fractal or multifractal structure. On the other hand, chaotic synchronization is a unique emergent self-organization phenomenon in nature. Classically, synchronization was characterized in terms of macroscopic parameters, such as the spectrum of Lyapunov exponents. Recently, however, we attempted a microscopic description of synchronization, called \textit{topological synchronization}, and showed that chaotic synchronization is, in fact, a continuous process that starts in low-density areas of the attractor.  Here we analyze the relation between the two emergent phenomena by shifting the descriptive level of topological synchronization to account for the multifractal nature of the visited attractors. Namely, we measure the generalized dimension of the system and monitor how it changes while increasing the coupling strength. We show that during the gradual process of topological adjustment in phase space, the multifractal structures of each strange attractor of the two coupled oscillators continuously converge, taking a similar form, until complete topological synchronization ensues. According to our results, chaotic synchronization has a specific trait in various systems, from continuous systems and discrete maps to high dimensional systems: synchronization initiates from the sparse areas of the attractor, and it creates what we termed as the 'zipper effect', a distinctive pattern in the multifractal structure of the system that reveals the microscopic buildup of the synchronization process. Topological synchronization offers, therefore, a more detailed microscopic description of chaotic synchronization and reveals new information about the process even in cases of high mismatch parameters.
\end{abstract}
\begin{document}

\flushbottom
\maketitle
\thispagestyle{empty}

\section*{Introduction}
Complex systems present us with an immense challenge as we try to explain their behavior. One key element in their description is how synchronization and self-organization emerge from systems that did not have these properties when isolated and, particularly, if the systems exhibit chaotic behavior. Synchronization underlies numerous collective phenomena observed in nature \cite{Pikovsky2001,Boccaletti2002,Strogatz2003,Boccaletti2018}, providing a scaffold for emergent behaviors, ranging from the acoustic unison of cricket choruses and the coordinated choreography of starling flocks \cite{Glass1988, Winfree2001} to human cognition, perception, memory and consciousness phenomena \cite{Ulhaas2006,Buzsaki,Bullmore2009,varela2001,Rodriguez1999,Klimesch1996,Singer2011}. Surprisingly, although chaotic systems have high sensitivity to initial conditions and thus defy synchrony, in the 1980's it has been shown that even chaotic systems can be synchronized \cite{pecora1990 ,pecora2015,Pikovskii1984,Huang2009}. Understanding how such a process can happen and characterizing the transition from completely different activities to synchrony in chaotic systems is fundamental to understanding the emergence of synchronization and self-organization in nature.

Chaotic dynamics present two fundamental and unique emergence phenomena, strange attractors which, in most cases, will have multifractal structure \cite{singularity ,Grassberger1983a,Grassberger1983b} and Chaotic synchronization. Understanding how these two phenomena occur and relate to each other is essential to shedding more light on the process of emergence in nature. Usually, chaotic synchronization is investigated by analyzing the time series of the system. Often it is observed by tracking the coordinated behavior of two slightly mismatched coupled chaotic systems, namely two systems featuring a minor shift in one of their parameters. As the coupling strength increases, a sequence of transitions occurs, beginning with no synchronization, advancing to phase synchronization \cite{jugphase}, lag synchronization \cite{bocavalladares}, and eventually, under sufficiently strong coupling, reaching complete synchronization. The process is typically characterized at the macroscopic level through the Lyapunov spectrum\cite{jugphase} and at the mesoscopic level through the non-localized unstable periodic orbits \cite{Pikovsky1997,Pazo2003,Yanchuk2003,Cvitanovic1991,Heagy1995}.

In  Ref.\cite{Lahav} we presented a new approach that revealed the microscopic level of the synchronization process. By presenting a new kind of synchronization, a \textit{topological synchronization}, we shifted descriptive levels of the synchronization process to the topology domain of the synced attractors. We discovered that synchronization is a continuous process at the microscopic level that starts from local synchronizations in different areas of the attractor. These local topological synchronizations start from the sparse areas of the attractor, where there are lower expansion rates, and accumulate until the system reaches complete synchronization. This paper investigates the relationship between the two emergent phenomena of chaos, the multifractal structure and strange attractors' synchronization process. In order to do so, we analyze the new phenomenon of topological synchronization. We show that topological synchronization of strange attractors is a gradual process at the emergent multifractal level. The multifractal structures of each strange attractor of the two coupled oscillators continuously converge to a  similar form until complete topological synchronization ensues. Topological synchronization unveils new detailed information about the synchronization process. Specifically, we characterize how the fractal dimensions change through the synchronization process and provide the probability of each scaling law to appear on the synchronized attractor and the probability of points to obey these scaling laws. In addition, after examination of various systems we show evidence that the chaotic synchronization process has a specific trait. Both in continuous system and discrete map and both in low and high dimensional systems, with the proper coupling, synchronization initiates from the sparse areas of the attractor and creates a \textit{zipper effect}. A distinctive pattern in the multifractal structure of the
system that reveals the microscopic buildup of the synchronization process. Lastly, we show that topological synchronization can also shade light in extreme synchronization cases between high mismatched coupled chaotic systems.

A multifractal structure typically characterizes the emergence of strange attractors \cite{Grassberger1983a,singularity}, which means that there is an infinite number of scaling laws in their structure, each captured by a different fractal dimension.
Furthermore, every scaling law has a different number of points that obeying it \cite{hentschel1983infinite}. \textit{Hausdorff dimension}, that typically captured by \textit{box count dimension}, is only one of these scaling laws.  
In order to demonstrate topological synchronization, we need to use a more general definition of dimension to account for this multifractality. To this end, we used \textit{R\' {e}nyi Generalized dimension} \cite{singularity,Renyi} which fully describes the structure of a multifractal for the different probabilities of each fractal:

\begin{equation}
\label{Dq}
D_q = \lim_{l \to 0} \left[ \frac{1}{q-1} \frac{\ln(\sum_i {p_i}^q)}{\ln(\frac{1}{l})}\right],
\end{equation}
where ${\mathrm{P}}_{\mathrm{i}}$ is the probability of a point (in state space) to be in sphere $i$, $l$ is the radius of the sphere and $q$ is a parameter that can be any real number. Parameter $q$ captures different fractal dimensions $D_q$ in the multifractal that have different probabilities for the trajectory to follow them. Thus, the general dimension is not just one value but a function of the parameter $q$. The dominant dimension is the box counting dimension, a mixture of all the scaling laws that will appear the most in the attractor, and it corresponds to $ \ D_0$. $\ D_1$ is the information dimension and$\ D_2$ is the correlation dimension \cite{D0}. $D_{-\infty }\ $ represents a  rare scaling law that appears only once in the strange attractor with a small probability of states obeying this law, and $D_{\infty }$ represents yet another very rare scaling rule that also appears once, but this time with a high probability of states obeying this law \cite{singularity,martinez1990clustering_Dq}.

\section*{Results}
Equipped with Eq.\ (\ref{Dq}) we can fully describe topological synchronization. We demonstrate it on one of the most fundamental examples in the context of synchronization, capturing two slightly mismatched chaotic R\"{o}ssler oscillators  \cite{roessler} coupled in a master-slave configuration. The equations of motion driving these oscillators take the form:

\begin{equation}\label{rossler}
  \begin{split}   
%\nonumber
\dot{\bf x}_1  &= f_{1}({\bf x}_1)
\\
\dot{\bf x}_2 &= f_{2}({\bf x}_2) +\sigmaup ({\bf x}_1-{\bf x}_2),
\end{split}
\end{equation}

\noindent
where ${\bf x}_1 \equiv (x_1,y_1,z_1)$ and ${\bf x}_2 \equiv (x_2,y_2,z_2)$ are the vector states of the master and slave oscillators respectively, $\sigmaup$ is the coupling strength and $f_{1,2}({\bf x})= (-y - z, x  + ay, b + z(x - c_{1,2}))$.
Without loss of generality we set the parameters to $a = 0.1$ and $b = 0.1$ identically across the two oscillators, and express the slight mismatch between the master and the slave through the parameters $c_1 = 18.0$ vs.\ $c_2 = 18.5$. System (\ref{rossler}) describes a unidirectional master (${\bf x}_1$) slave (${\bf x}_2$) form of coupling, uniformly applied to all coordinates $x,y$ and $z$. Under this directional coupling scheme, we can track and quantify the process of synchronization in a controlled fashion, as the slave gradually emulates the behavior of the master while the master continues its undisturbed oscillations.

\begin{figure}
\centering
\includegraphics[width=4in]{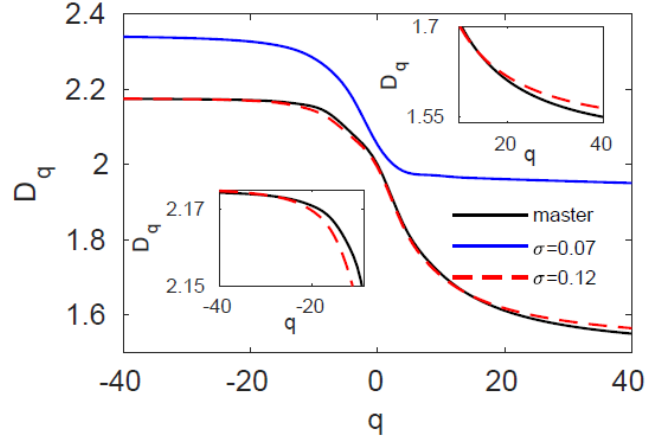}
%\vspace{-4mm}
\caption[]{(Color online). \textbf{Generalized fractal dimension for slightly mismatched R\"{o}ssler systems}. General dimension $D_q$ as a function of parameter $q$ for the master (black) and slave for coupling strengths $\sigma = 0.07$ (blue solid) and $\sigma = 0.12$ (red dashed). Insets are blow-ups for the master and slave curves for $\sigma = 0.12$ in the $q< 0$ (bottom left) and $q>0$ (top right) regions. Topological synchronization occurs as the $D_q$ curve of the slave matches the $D_q$ curve of the master.}
\label{image1}
%\vspace{-3mm}
\end{figure}

In Ref.\cite{Lahav},  we showed the microscopic buildup of synchronization in the system (\ref{rossler}). Local synchronization initiates in the sparse areas of the attractor and as the local synchronizations accumulate, phase synchronization occurs for $\mathrm{\sigmaup}_{ps}\mathrm{\ge }\mathrm{0.1\ }$ and complete synchronization is obtained for $\mathrm{\sigmaup_{cs}\ }\mathrm{\ge }\mathrm{2.0}$. In  Fig.\ \ref{image1} we show the general dimension curves, $D_q$ of the system (\ref{rossler}), which describes the process of the topological synchronization between the master and the slave. The master (black) has a fixed curve while the slave starts with a completely different $D_q$ than the master in low coupling $\sigmaup$ = 0.07 (blue) and converges with the master at higher coupling $\sigmaup$ = 0.12 (red dashed). Moreover, at the transition point to phase synchronization (red dashed), blow-ups for the master and slave curves show that in the slave case, $D_q$ for $q\mathrm{<} 0$ is much closer to the master than $D_q$ in the positive part ($q \mathrm{>} 0$). If we compare the two zooms on the positive and~negative parts of the curve and consider the difference in the vertical axis ranges for negative ($0.02$) and positive ($0.15$) $q$, we observe a difference of almost one order of magnitude. This result corresponds to the fact that local synchronization initiates in the sparse areas of the attractor where the probability of points is low. As we move to the negative part of the parameter $q$ and approach $D_{-\infty }\ $, we examine the sparse areas of the attractor with low probability scaling laws, and indeed they reach topological synchronization before the dense areas of the attractor (for $q>0$).

The previous example demonstrates that the synchronization process between different strange attractors can be understood as a topological synchronization between the multifractal structures of the attractors. Topological synchronization means that the multifractal structure of one attractor predicts the multifractal structure of the second attractor, and when complete topological synchronization occurs, the multifractal structure of one attractor fully predicts the other. Thus, topological synchronization is characterized by the boundedness of the difference between the $D_q$ curves of the two oscillators over the whole dynamical evolution of the system. Consequently, the condition for complete topological synchronization between oscillator 1 and 2 is:

\begin{equation}
\label{toposinc}
\Delta D_q=\sup_q |{D_q}^{(1)}-{D_q}^{(2)}|\to 0.
\end{equation}

\begin{figure}
\centering
\includegraphics[width=2.5in]{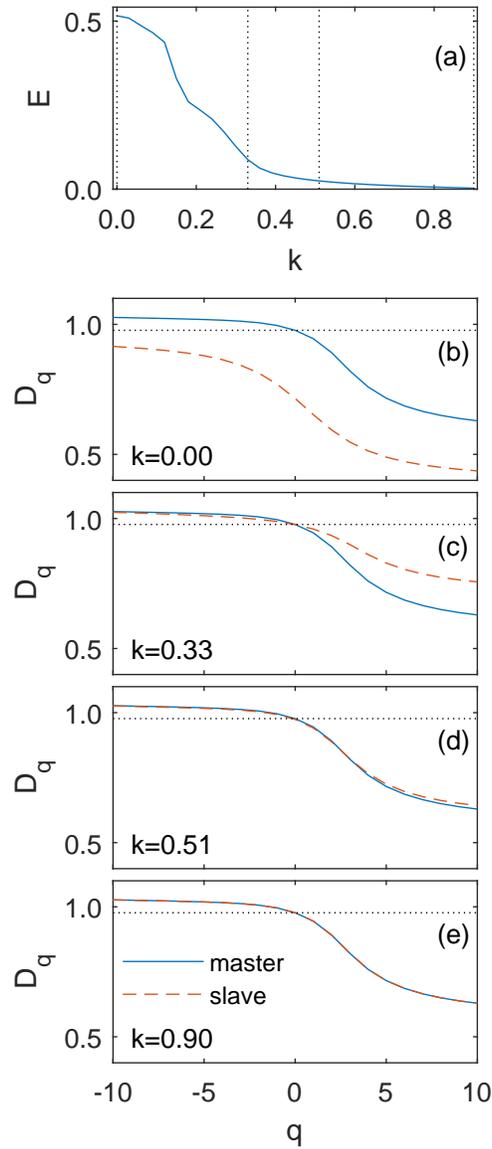}
%\vspace{-4mm}
\caption[]{(Color online). \textbf{Microscopic build-up of synchronization for the Logistic map system}.  (a) Synchronization error $E$ as function of coupling strength \textit{k}. Complete synchronization $E\sim 0$ is reached around $k\sim 0.9$. (b)-(e) Topological synchronization and the zipper effect. General dimension $D_q$ as function of the parameter $q$ of master (blue) and slave (red) attractors. As the coupling $k$ increases a zipper effect from the negative  ($q\le 0$) to the positive  ($q>0$) part of $D_q$ is seen.}
\label{image2}
%\vspace{-3mm}
\end{figure}

\begin{figure}
\centering
\includegraphics[width=2.5in]{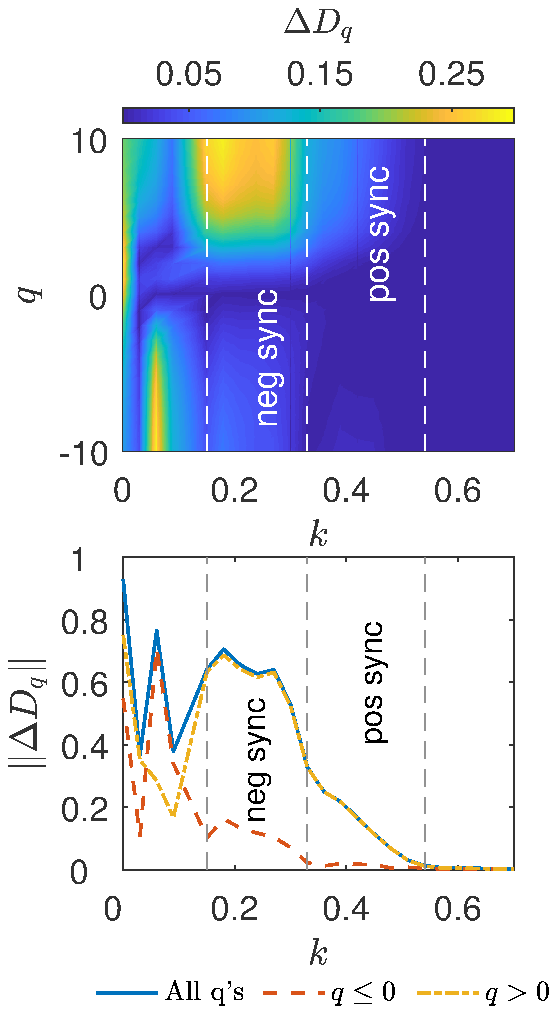}
%\vspace{-4mm}
\caption[]{(Color online). \textbf{Distance between the master and slave's general dimensions for the mismatched Logistic map systems}. Upper panel, color map denoting the distance between the  $D_q$ curves of the master and the slave, $\Delta D_q$ as a function of the parameter $q$ (y axis) and the coupling strength $k$ (x axis). Vertical dashed lines show the negative and positive zipper effect regions (as $\Delta D_q$ goes to zero). Bottom panel, distance between the  master and slave $D_q$ curves calculated as the norm between them,  $||\Delta D_q||$ as function of $k$ in the $q\le 0$ (red dashed curve), $q>0$ (yellow dot-dashed curve) intervals and in the whole range of $q$ (blue solid line). At $k\sim 0.33$, $D_{q<0}$ of the slave has completed its synchronization with the master and its $D_{q>0}$ starts a gradual approach to the master curve. The zipper effect is completed around $k\sim 0.9$. Vertical dashed lines mark the negative and positive zipper effect regions.}
\label{image3}
%\vspace{-3mm}
\end{figure}

In order to further analyze the properties of topological synchronization we chose a simple 1D discrete system from the Logistic map family, coupled in a master-slave configuration. The equations of motion driving these oscillators take the form \cite{grassberger1981LogisticMap}:

\begin{align}
\nonumber
\label{logistic}
& x_{n+1}=\mathrm{\ }{\mathrm{c}}_1\left(1-2x^2_n\right)
\\
& y_{n+1}=\ (1-k){\mathrm{c}}_2\left(1-2y^2_n\right)+{\mathrm{c}}_1k(1-2x^2_n),
\end{align}
\noindent
where $k$ is the coupling strength. Without loss of generality, we express the mismatch between master and  slave through the parameters $c_{1}= 0.89$ and $c_{2} = 0.8373351$, respectively. The slave oscillator ($y_n$) is on the onset of chaos with sparse strange attractor whereas the master  ($x_n$) has a dense strange attractor. In Fig.\ \ref{image2}a we present the synchronization error parameter $E$ versus $k$ calculated as $E =  \tau^{-1} \int_{t_0}^{t_0+\tau} \|{\bf x}_1-{\bf x}_2\| dt$ (letting the system to evolve from random initial conditions with fixed integration time step of $h=0.001$ from $t=0$ up to $t_0=50,000$ t.u. and averaging the distance between the oscillators' states during $\tau=1,000$ t.u.).  As $E\mathrm{\to}0$ at $k_{CS}\mathrm{\sim } 0.9$ complete synchronization emerges. Topological synchronization unveils the microscopic process underlying synchronization.
The microscopic buildup is caused by a topological matching mechanism which eventually leads to complete synchronization between the two attractors.  Figures\ \ref{image2}b-e and  \ref{image3} examine the general dimension of the system (\ref{logistic}) and reveal this topological synchronization process. Both figures show that a gradual increase of $k$ causes a gradual decrease of the distance between the two $D_q$ curves to zero. According to  Fig.\ \ref{image3}, around $k=0.21$ the distance between the $D_q$ curves for $q\le 0$ begins to decrease until it vanishes around $k=0.33$, whereas in the region $q>0$ the distance begins to decrease at only around $k=0.3$. The system reaches complete topological synchronization with zero distance between the two $D_q$ curves at $k=0.9$.
%The synchronization points percentage (\textit{SPP}) index, quantifies the fraction of points in phase space for which there exists a local continuous surjective function to map the state of the master with that of the slave. If, within some area in the vicinity of these points such a continuous mapping exists, one can predict the specific state of the slave system directly from measuring the master state, representing a local topological coherence of the master-slave duo \cite{boccapeco,Pastur04,Pecora95}. Thus, the \textit{SPP} index can monitor the gradual passage from local to global synchronization in the system.

Furthermore, in  Fig.\ \ref{image2}b-e, we show that the changes of the slave $D_q$ curve versus $k$ reveal a \textbf{zipper effect} of the general dimension from the negative $q$ to the positive $q$. At low coupling, there is a continuous synchronization of the negative part of the $D_q$ curve ($\mathrm{q}\mathrm{\le }\mathrm{0}$). When the negative part of the $D_q$ curve is synchronized around $k=0.33$ (panel c), the positive part starts to synchronize. More specifically, $D_{1}$ synchronizes at around $k=0.36$, $D_{2}$ at around $k=0.42$, and $D_{3}$ at ${k}=0.51$ (panel d) and so on, ``zipping'' the topological synchronization process until at around $k=0.9$, where complete synchronization is achieved, the $D_{10}$ dimension of the slave equals that of the master (panel e. For video of the whole zipper effect process, see \href{https://www.dropbox.com/s/riu0t0ycfw73t80/movie_Logistic_map_Dq_Strong_master_Weak_slave_k0_0002_k0pt9.mp4?dl=0}{supplementary video 1} and \href{https://www.dropbox.com/s/p61wdztxjzteyrt/Logistic_map_attractorHist_strong_master_week_slave_k0_k0pt6_distance.mp4?dl=0}{supplementary video 2}).

The finding of negative to positive zipper effect in the $D_q$ curves concurs also with the  R\"{o}ssler system. As noted before, stepping from $D_{-\infty }\ $to $D_{\infty }\ $ represents stepping from scaling laws with low occupation probability to scaling laws with high occupation probability. The zipper effect, thus, implies that as in the R\"{o}ssler case, also in the Logistic map case, topological synchronization starts at low coupling strength at areas of the attractor that have a low probability of points and, only when these areas complete their local synchronization,  the attractor will topologically synchronize also in areas with a high probability of points at a large coupling.
%Fig.\ref{image4}b. presents the distribution of local Synchronization points, $\sigma$ (normalized \textit{SPP}) on the attractor and comperes them to the attractor density, $\rho$. Indeed, as suspected, local synchronization initiates in the sparse regions of the attractor in agreement with the topological zipper effect.

\begin{figure}
\centering
\includegraphics[width=2.5in]{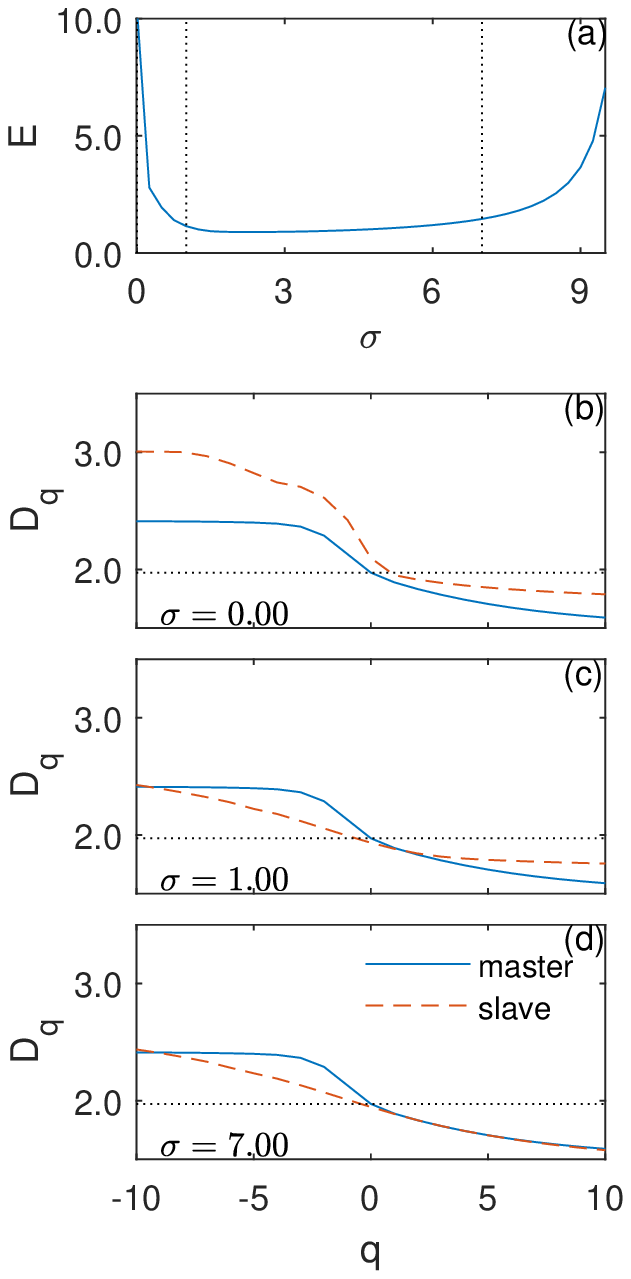}
%\vspace{-4mm}
\caption[]{(Color online). \textbf{ Microscopic build-up of synchronization for high mismatched R\"{o}ssler systems}. (a) Synchronization error $E$ as function of coupling strength $\sigmaup$.  We obtain a window of approximate complete synchronization, in which $E \ll 1$, between $\sigmaup^{1}_{\rm CS} \sim 3$ and $\sigmaup^{2}_{\rm CS} \sim 7$. (b)-(d) Topological synchronization and the zipper effect. General dimension $D_q$ as function of the parameter $q$ of master (blue) and slave (red) attractors. As coupling $\sigmaup$ increases a zipper effect from the negative to the positive part of $D_q$ can be seen. (c) At low couplings, the slaves' negative $D_q$ part syncs and reaches a minimal stable distance from the master. (d) The positive $D_q$ part synchronizes only after the negative part has completed to synchronize and reaches a stable minimal distance from the master as well.}
\label{image4}
%\vspace{-3mm}
\end{figure}

\begin{figure}
\centering
%\vspace{-4mm}
\includegraphics[width=2.5in]{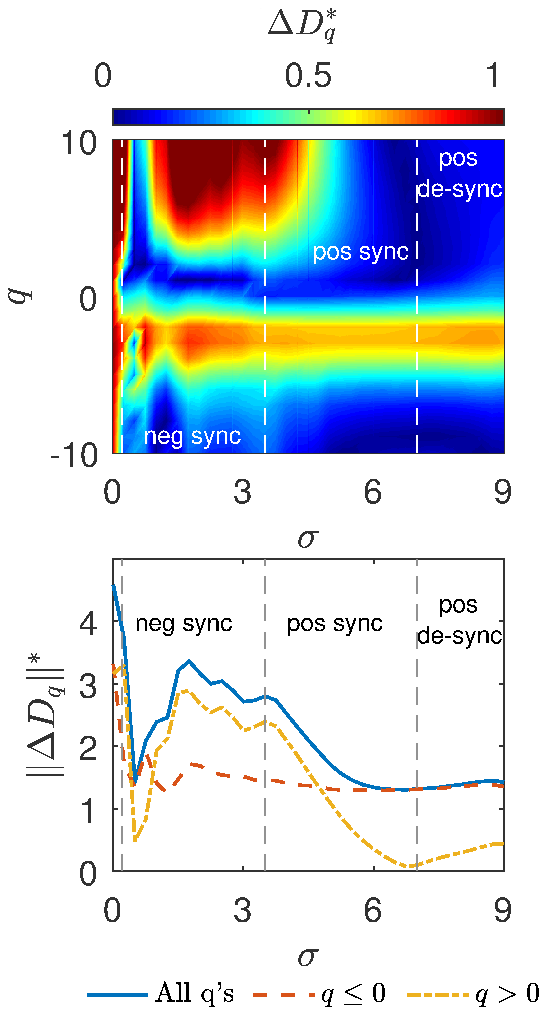}
\caption[]{(Color online). \textbf{Distance between the general dimension of master and slave for high mismatched R\"{o}ssler systems}. Upper panel, color map denoting the normalized distance between the  $D_q$ curves of the master and the slave, $\Delta D_q^*$  as a function of the parameter $q$ (y axis) and the coupling strength $\sigmaup$ (x axis). Vertical dashed lines show the negative and positive zipper effect regions (as distance decreasing to a minimum fixed value). Bottom panel, the normalized distance between the  $D_q$ curves of the master and the slave calculated as the norm between them,  $||\Delta D_q^*||$ as a function of $\sigmaup$ in the  $q\le 0$ (red dashed curve), $q>0$ (yellow dot-dashed curve) intervals and in the whole range of $q$ (solid blue line). At  $\sigmaup\sim 3$, the negative part of $D_q$ has completed its synchronization, and the positive part starts a gradual approach to the master curve. The zipper effect completes around $\sigmaup\sim 6$. At $\sigmaup\sim 7$ the system exits the synchronization window with a reverse zipper effect. The positive part of $D_q$  gradually separates from the master curve (and thus the distance from the master increases) while the negative part of $D_q$ remains with the same close distance to the master. Vertical dashed lines show the negative and positive zipper effect regions.}
\label{image5}
%\vspace{-4mm}
\end{figure}

In both discrete map and continuous systems, we see the same distinctive pattern of the zipper effect in the multifractal structure, where topological synchronization starts from the sparse to the dense areas of the attractor, suggesting that this trait might be an essential feature in chaotic synchronization. 
%If a system can reach chaotic synchronization, one can find a coupling form in which the synchronization initiates from the sparse areas of the attractor and creates a zipper effect. Namely, from synchronization of sparse areas in the attractor to synchronizations of increasingly more crowded areas in the attractor until only with sufficient coupling strength a global complete topological synchronization can be achieved. 

In order to validate the robustness of the zipper effect,
% is indeed a characteristic trait of chaotic synchronization, 
we analyzed the topological synchronization of several systems with different features, a high mismatch class III system and a high dimensional system. For the first system, we chose a special case of the R\"{o}ssler system with a high mismatch between the parameters of the system and with a different form of the master-slave coupling by taking $c_1=14$ for the master and $c_2=18$ for the slave coupled through the variable $x$. This can be achieved by substituting the coupling term in Eq.\ (\ref{rossler}) with $\sigmaup(x_1 - x_2)$, a coupling applied uniquely to the $x$ variable. This change turns the coupled oscillators into a class III system, in which synchronization is confined to a finite interval in $\sigmaup$ \cite{Boccaletti2006,Huang2009}. Indeed, now the system features a first transition at $\sigmaup = \sigmaup^1_{\rm CS}$, in which it enters almost complete synchronization, followed by a second transition at $\sigmaup = \sigmaup^2_{\rm CS} > \sigmaup^1_{\rm CS}$, in which it begins to de-synchronize (Fig.\ \ref{image4}a).
% We express the mismatch between the master and the salve through the parameters c${}_{1}$ = 14 (master) versus c${}_{2}$ = 18 (slave). 

In such an extreme scenario, the multifractal structures of the two oscillators are sufficiently different to observe the process of the two distinctive $D_q$ curves converge. Because of the system's high mismatch, a much higher coupling strength is needed to synchronize, but complete topological synchronization can not be achieved. Nevertheless, an approximate zipper effect occurs here as well. 
% We used equation similar to system (\ref{rossler}), but this time the unidirectional master (\textbf{x${}_{1}$}) slave (\textbf{x${}_{2}$}) form is coupled only in the x coordinate. We express the mismatch between the master and the salve through the parameters c${}_{1}$ = 18 (master) versus c${}_{2}$ = 14 (slave).

Figure \ref{image4} presents the microscopic buildup of synchronization for the high mismatched R\"{o}ssler systems. The synchronization error parameter $E$ approaches zero at a window of approximate complete synchronization, in which $E \ll1$, between $\sigmaup^{1}_{\rm CS} \sim 3$ and $\sigmaup^{2}_{\rm CS} \sim 7$ (Fig.\ref{image4}a).  Figure \ref{image4}b-d reveals that indeed, there is a zipper effect of the general dimension from the negative part of $D_q$ ($q\le 0$) to the positive part of $D_q$  ($q>0$) also in the high mismatch R\"{o}ssler system. The slaves' negative $D_q$ part synchronizes at low coupling and reaches a minimal stable distance from the master (panel c). The positive $D_q$ part start its' synchronization only after the negative part finished to sync around  $\sigmaup \mathrm{\sim }$3.5 and the system approaches an almost complete synchronization at $\sigmaup \mathrm{\sim }6$ (panel d. For video of the whole zipper effect process, see \href{https://www.dropbox.com/s/7sil2bwf6e7fvdj/rossler_Dq_14m_18s_coupx_N100k_Nc8000_v4.mp4?dl=0}{supplementary video 3} and \href{https://www.dropbox.com/s/euogijgragob0ni/Rossler_attractor_14m_18s_coupx_N100k_Nc8000_v4.mp4?dl=0}{supplementary video 4}). Figure  \ref{image5} shows $|\Delta D_q|^*$, the distance between the  $D_q$ curves of the master and the slave normalized by  the initial distance between the  $D_q$ curves at $\sigmaup=0$. As the coupling strength $\sigmaup$ increases, a gradual decrease of the distance between the two $D_q$ curves is observed. Around  $\sigmaup \mathrm{\sim }0.25$ the distance of the negative part of the  $D_q$ curve begins to decrease until it reaches a minimal stable distance from the master around  $\sigmaup \mathrm{\sim }3$ whereas the distance of the positive part of the $D_q$ curves begins to decrease only at around $\sigmaup \mathrm{\sim }3.5$ and reaches zero at around $\sigmaup \mathrm{\sim }6$.  Interestingly, both in the Logistic map case and in high mismatched  R\"{o}ssler system the first dimension that syncs with the master and starts the zipper effect in the $q\le 0$ is $D_{0}$ (which is a mixture of all the scaling laws that will appear the most in the attractor  \cite{singularity}. See supplementary material note 1). Above $\sigmaup^2_{\rm CS}\mathrm{\sim }7$, the system exits the synchronization window with a reverse zipper effect. The positive part of $D_q$  gradually increases its distance from the master curve while the negative part of $D_q$ remains with the same distance from the master. The physical interpretation of the reverse zipper effect is that as the system exits the synchronization window, the dense regions of the attractor will de-synchronize before the low density regions. Indeed, in \cite{Lahav}, we showed that in class III systems, de-synchronization starts from the dense regions of the attractor. 

\begin{figure}
\centering
%\vspace{-4mm}
\includegraphics[width=2.5in]{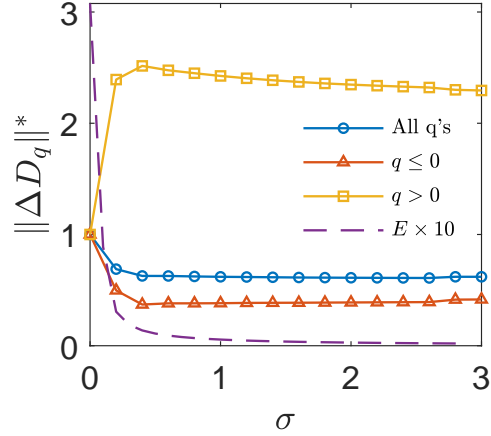}
\caption[]{(Color online). \textbf{Distance between the general dimension of master and slave for high dimensional Mackey-Glass system}. The normalized distance between the  $D_q$ curves of the master and the slave calculated as the norm between them,  $||\Delta D_q^*||$ as a function of $\sigmaup$ in the  $q\le 0$ (red triangles), $q>0$ (yellow rectangles) intervals and in the whole range of $q$ (blue circles) and the synchronization error parameter $E$ multiplied by 10 (purple dashed line). At  $\sigmaup\sim 0.5$, the negative part of $D_q$ has completed its synchronization, and the positive part starts to sync.}
\label{image6}
%\vspace{-4mm}
\end{figure}

For the second system we chose the Mackey-Glass equation \cite{mg}. This equation contains a delay $\tau$, which makes it an infinite dimensional system. According to Farmer \cite{farmer}, as the value of $\tau$ is increased, the dimension of the attractor increases as well:
\begin{align}
\nonumber
\label{mg}
&\dot{x}_1(t) = -\gamma x_1(t) + \beta\frac{x_1(t-\tau_1)}{1+x_1(t-\tau_1)^{n}}
\\
&\dot{x}_2(t) = -\gamma x_2(t) + \beta\frac{x_2(t-\tau_2)}{1+x_2(t-\tau_2)^{n}}+\sigmaup\left[ x_1(t) - x_2(t)\right],
\end{align}
\noindent
where $\sigmaup$ is the coupling strength and $\tau_1$, $\tau_2$ are the delays of oscillator 1 and 2 respectively. Without loss of generality we set the parameters to $\gamma = 0.1$, $\beta = 0.2$ and $n=10$ identically across the two oscillators. We express the mismatch between the master and the slave through the delays, $\tau_1=27$ and $\tau_2=26$. With such delays we could explore topological synchronization in high dimensional space. In order to explore the attractors generating from Eq.\ (\ref{mg}) we need to use a time delay embedding of the equations, with embedding delay of $\tauup^\prime=32$ and embedding dimension of $D= 6$. Figure \ref{image6} presents both the synchronization error parameter $E$  and $|\Delta D_q|^*$, the distance between the $D_q$ curves of the master and the slave normalized by the initial distance between the $D_q$ curves at $\sigmaup=0$. As the coupling strength $\sigmaup$ increases the synchronization error parameter $E$ approaches zero and a gradual decrease of the distance between the two $D_q$ curves is observed. In low coupling strengths the distance of the negative part of the  $D_q$ decreases until it reaches a minimal stable distance from the master around  $\sigmaup \mathrm{\sim }0.5$ whereas the distance of the positive part of the $D_q$ curves begins to decrease only after $\sigmaup \mathrm{\sim }0.5$ to reach a minimal stable distance from the master as well. as a result, there is a zipper effect of the general dimension from the negative part of $D_q$ ($q\le 0$) to the positive part of $D_q$  ($q>0$) also in high dimensional Mackey-Glass system. The positive $D_q$ part start to synchronize only after the negative part finished to sync and the system approaches an almost complete synchronization at $\sigmaup \mathrm{\sim }2$.

\section*{Discussion}

In this paper, we analyzed the relationship between the emergence of two phenomena of chaotic dynamics. On the one hand, the multifractal structure of a strange attractor and, on the other hand,  chaotic synchronization. We demonstrate this relationship by introducing topological synchronization, in which the multifractal structure of one strange attractor approaches the other until the multifractal structure of the attractors is the same. Topological synchronization shifts the descriptive levels of synchronization to the emergence level of the topology domain of the attractors. Topological synchronization is a powerful tool to investigate chaotic synchronization, and it gives us relevant information even in high mismatched systems and in high dimensional systems. It reveals that chaotic synchronization is a continuous process that can be described by the zipper effect. The fact that we see the same zipper effect in various different systems, from logistic map and the canonical  R\"{o}ssler system to an extreme case of high mismatched  R\"{o}ssler system and high dimensional chaotic system like the Mackey-Glass system, supports that the zipper effect is an essential trait of chaotic synchronization. Furthermore, we even see that in class III system, not only that the zipper effect is present during the synchronization process, it is also present during the de-synchronization process as a reverse zipper effect. Our findings, therefore, suggest that, typically, the road to complete synchronization starts at low coupling with topological synchronization of the sparse areas in the attractor and continues with topological synchronizations of much more dense areas in the attractor until complete topological synchronization is reached for high enough coupling.
The study of chaotic microscopic behavior can only be observed through numerical tools, and motivated and supported, as we did in the paper, by reasoning and qualitative arguments. As a result, in order to build a solid case, topological synchronization needs to be studied more and additional numerical studies need to be performed in order to continue and validate the results that we presented in this paper.

One application of these results is to determine how much synchronization a physical system has and where, in phase space, it occurred. For some real chaotic systems, complete synchronization will be detected, whereas other systems may only sync until the point where their less crowded areas in the attractor will be synchronized. Topological synchronization can detect these differences and show which areas of the phase space have already synchronized. Furthermore, by applying topological synchronization to extreme cases of chaotic systems (like cases of high mismatch between the parameters of the system) we can extend our understanding of the synchronization process to cases that were hard to study before.

\section*{Acknowledgements}

The authors would like to thank Ashok Vaish  for his continuous support.
C.H. is supported by INSPIRE-Faculty grant (Code:IFA17-PH193).
This work was supported by the US National Science Foundation - CRISP Award Number: 1735505 and by the Ministerio de Econom\'ia y Competitividad of Spain (project FIS2017-84151-P) and Ministerio de Ciencia e Innovaci\'on (project PID2020-113737GB-I00).

 \section*{Author contributions statement}
N.L. conception of the work, C.H., B.K., I.S.N. coding, N.L., C.H., B.K. data simulation aquasition and analysis, N.L. Drafting the article. All authors reviewed the manuscript.


\begin{thebibliography}{99}
\bibitem{Pikovsky2001}
  A. Pikvosky, M. Rosenblum,  and J.  Kurths,  {\it Synchronization: a Universal Concept in Nonlinear Sciences} (Cambridge University Press, Cambridge, England, 2001).
\bibitem{Boccaletti2002} S. Boccaletti, J. Kurths, G. Osipov,  D. L. Valladares, and  C. S. Zhou,  Phys.~Rep. {\bf366}, 1 (2002).
  \bibitem{Strogatz2003}
    S. Strogatz,  {\it Sync: The emerging science of spontaneous order} (Hyperion, 2003).
    \bibitem{Boccaletti2018}
S. Boccaletti, A. Pisarchik, C.I. del Genio and A. Amann {\it Synchronization: from Coupled Systems to Complex Networks} (Cambridge University Press, Cambridge, England, 2018).

\bibitem{Glass1988}
  L. Glass and M. C. Mackey,  {\it From clocks to chaos: the rhythms of life} (Princeton University Press, 1988).
  \bibitem{Winfree2001}
A. T. Winfree, {\it The geometry of biological time, Vol. 12} (Springer Science \& Business Media, 2001).

\bibitem{Ulhaas2006}
P. J. Uhlhaas and W. Singer, {\it Neuron} {\bf 52}, 155 (2006).
\bibitem{Buzsaki} G. Buzsaki,  {\it Rhythms of the Brain} (Oxford University Press, 2006).
  \bibitem{Bullmore2009} E. Bullmore and  O. Sporns,  Nat.~Rev.~Neurosc. {\bf 10}, 186 (2009).

\bibitem{varela2001}
Varela, Francisco and Lachaux, Jean-Philippe and Rodriguez, Eugenio and Martinerie, Jacques, {\it Nature reviews neuroscience} {\bf 2}, 4 (2001).

\bibitem{Rodriguez1999}
E. Rodriguez, N. George, J.-P. Lachaux, J. Martinerie, B. Renault and F. J. Varela,  {\it Nature} {\bf 397}, 430 (1999).

\bibitem{Klimesch1996}
W. Klimesch, {\it International Journal of Psychophysiology} {\bf 24}, 61 (1996).

\bibitem{Singer2011}
W. Singer,  {\it Consciousness and neuronal synchronization (in: The neurology of consciousness)} (Academic Press, 2011).

\bibitem{pecora1990}
L.M. Pecora and T.L. Carroll, Phys. Rev. Lett. {\bf 64}, 821 (1990).

\bibitem{pecora2015}
L.M. Pecora and T.L. Carroll, Chaos {\bf25}, 097611 (2015).

\bibitem{Pikovskii1984}
A. Pikovskii,  Z. Phys. B {\bf55}(2), 149 (1984).

\bibitem{Huang2009}
L Huang, Q Chen, YC Lai, LM Pecora, Phys. Rev. E, \textbf{80}, 036204 (2009)

\bibitem{singularity}
T. C. Halsey, M. H. Jensen, L. P. Kadanoff, I. Procaccia, B. I. Shraiman, Physical Review A \textbf{33}.2, 1141 (1986).

\bibitem{Grassberger1983a}
P Grassberger, I Procaccia,  Physica D \textbf{9}, 189-208 (1983).
\bibitem{Grassberger1983b} P Grassberger, I Procaccia,  Phys. Rev. Let. \textbf{50}, 346-349 (1983).

  \bibitem{jugphase}
    M.G. Rosenblum, A.S. Pikovsky, and J. Kurths, Phys. Rev. Lett. {\bf 76}, 1804 (1996).
    \bibitem{bocavalladares}
      S. Boccaletti and D. L. Valladares, Phys. Rev. {\bf E62}, 7497 (2000).
      \bibitem{Pikovsky1997}
A. Pikovsky, M. Zaks, M. Rosenblum, G. Osipov and J. Kurths, Chaos \textbf{7}, 680-687 (1997).

\bibitem{Pazo2003}
D. Pazo, M. Zaks and J. Kurths, Chaos, \textbf{13}, 309-318 (2003).

\bibitem{Yanchuk2003}
S. Yanchuk, Y. Maistrenko and E. Mosekilde, Chaos, \textbf{13}, 388-400 (2003).

\bibitem{Cvitanovic1991}
P. Cvitanović, Physica D, \textbf{51}, 138-151 (1991).

\bibitem{Heagy1995}
  Heagy, J. F., T. L. Carroll, and L. M. Pecora, Physical Review E \textbf{52}, R1253 (1995).

  
\bibitem{Lahav}
N. Lahav, I. Sendi\~na-Nadal, C. Hens, B.
Ksherim, B. Barzel, R. Cohen, S. Boccaletti,  Physical Review E \textbf{98}, 052204 (2018).
\bibitem{hentschel1983infinite}
Hentschel, HGE and I. Procaccia, Physica D: Nonlinear Phenomena \textbf {8}, 435--444 (1983).

\bibitem{Renyi}
A.  Renyi,  Dimension,  entropy  and information,    Transactions  of the Second  Prague Conference  on Information  Theory,  Academic  Press,  New York,  1960,  545-556.

\bibitem{D0}
P. Grassberger, I. Procaccia, Physica D, \textbf{13}.1-2, 34-54 (1984).

\bibitem{martinez1990clustering_Dq}
Martinez V.j., Jones B. J.T., Dominguez-Tenreiro R., Van de Weygaert R., Astrophysical Journal,  \textbf{357}, 50 (1990).

%\bibitem{boccaexp}
%S. Boccaletti, E. Allaria, R. Meucci and F.T. Arecchi, Phys. Rev. Lett. {\bf 89}, 194101 (2002).

\bibitem{roessler}
O. E. R\"ossler, Physics Letters A {\bf 57}, 397-398 (1976).

\bibitem{grassberger1981LogisticMap}
P. Grassberger, Journal of Statistical Physics {\bf 26}, 173--179 (1981).


% \bibitem{boccapeco}
% S. Boccaletti, L.M. Pecora and A. Pelaez, Phys. Rev. {\bf E63}, 066219 (2001).

% \bibitem{Pastur04}
% L. Pastur, S. Boccaletti, and P. L. Ramazza, Phys. Rev. E {\bf 69},  036201 (2004).

% \bibitem{Pecora95}
% L. M. Pecora, T. L. Carroll, and J. F. Heagy, Phys. Rev. E {\bf 52},
% 3420 (1995).

\bibitem{Boccaletti2006}
S. Boccaletti, V. Latora, Y. Moreno, M. Chavez, and D.-U. Hwang, Phys. Rep., \textbf{424}, 175-308 (2006).

% \bibitem{Huang2009}
% L Huang, Q Chen, YC Lai, LM Pecora, Phys. Rev. E,\textbf{80}, 036204 (2009)

\bibitem{mg}
MC Mackey, L Glass, Science,\textbf{197}, 4300 (1977).

\bibitem{farmer}
D Farmer, Physica D,\textbf{4}, 3 (1982).

\end{thebibliography}
\end{document}